\documentclass[10pt,journal,compsoc]{IEEEtran}
\usepackage[dvips]{graphicx}
\usepackage{epsfig,multirow}
\usepackage{amsmath,amssymb,bm}
\usepackage{amsfonts,dsfont,color,bbm}
\usepackage[linesnumbered,lined,commentsnumbered,boxed,ruled]{algorithm2e}
\usepackage{caption}
\usepackage{subcaption}
\usepackage{epstopdf,epsf,epsfig}
\usepackage[nocompress]{cite}
\usepackage{mathtools}
\usepackage{float}
\usepackage{graphicx}
\usepackage{pgfplots}
\usepackage{tikz}
\usetikzlibrary{external}

\renewcommand{\vec}[1]{\mbox{\boldmath${#1}$}}

\DeclareMathOperator*{\argmin}{arg\,min}

\usepackage{fancyhdr}

\title{Efficient Implementation Of Newton-Raphson Methods For Sequential Data Prediction}

\author{Burak C. Civek and Suleyman S. Kozat, \textit{Senior Member, IEEE}
	\IEEEcompsocitemizethanks{ \IEEEcompsocthanksitem  Burak C. Civek and S. S. Kozat are with the Department of Electrical and Electronics Engineering, Bilkent University, Ankara 06800, Turkey (e-mail: \{burak,kozat\}@ee.bilkent.edu.tr).}
\thanks{This work is supported in part by Turkish Academy of Sciences Outstanding Researcher Programme, TUBITAK Contract No. 113E517.}}

\begin{document}
	\IEEEtitleabstractindextext{
		\begin{abstract}
			We investigate the problem of sequential linear data prediction for real life big data applications. The second order algorithms, i.e., Newton-Raphson Methods, asymptotically achieve the performance of the "best" possible linear data predictor much faster compared to the first order algorithms, e.g., Online Gradient Descent. However, implementation of these methods is not usually feasible in big data applications because of the extremely high computational needs. Regular implementation of the Newton-Raphson Methods requires a computational complexity in the order of $O(M^2)$ for an $M$ dimensional feature vector, while the first order algorithms need only $O(M)$. To this end, in order to eliminate this gap, we introduce a highly efficient implementation reducing the computational complexity of the Newton-Raphson Methods from quadratic to linear scale. The presented algorithm provides the well-known merits of the second order methods while offering the computational complexity of $O(M)$. We utilize the shifted nature of the consecutive feature vectors and do not rely on any statistical assumptions. Therefore, both regular and fast implementations achieve the same performance in the sense of mean square error. We demonstrate the computational efficiency of our algorithm on real life sequential big datasets. We also illustrate that the presented algorithm is numerically stable.
		\end{abstract}
		}
\maketitle
\IEEEdisplaynontitleabstractindextext

\section{Introduction}\label{sec:intro}
\IEEEPARstart{T}{echnological} developments in recent years have substantially increased the amount of data gathered from real life systems\cite{mining,TKDE1,bigdata2,bigdata1}. There exists a significant data flow through the recently arising applications such as large-scale sensor networks, information sensing mobile devices and web based social networks \cite{TKDE2,bigdata3,sensor2}. The size as well as the dimensionality of this data strain the limits of current architectures. Since processing and storing such massive amount of data result in an excessive computational cost, efficient machine learning and data processing algorithms are needed \cite{mining,Moon1}. \par
In this paper, we investigate the widely studied sequential prediction problem for high dimensional data streams. Efficient prediction algorithms specific to big data sequences have great importance for several real life applications such as high frequency trading\cite{HFT}, forecasting\cite{financial}, trend analysis\cite{trend}, financial market \cite{financial2} and locational tracking\cite{location}. Unfortunately, conventional methods in machine learning and data processing literatures are inadequate to efficiently and effectively process high dimensional data sequences \cite{Bottou,Bottou2,cesa_book}. Even though today's computers have powerful processing units, traditional algorithms create a bottleneck even for that processing power when the data is acquired at high speeds and too large in size \cite{Bottou,Bottou2}.           
\par  
In order to mitigate the problem of excessive computational cost, we introduce sequential, i.e., online, processing, where the data is neither stored nor reused, and avoid "batch" processing. \cite{singer,cesa_book}. One family of the well known online learning algorithms in the data processing literature is the family of first order methods, e.g., Online Gradient Descent \cite{Hazan,optimization}. These methods only use the gradient information to minimize the overall prediction cost. They achieve logarithmic regret bounds that are theoretically guaranteed to hold under certain assumptions \cite{Hazan}. Gradient based methods are computationally more efficient compared to other families of online learning algorithms, i.e., for a sequence of $M$-dimensional feature vectors $\{\vec{x}_t\}_{t \geq 0}$, where $\vec{x}_t \in \mathbbm{R}^M$, the computational complexity is only in the order of $O(M)$. However, their convergence rates remain significantly slow when achieving an optimal solution, since no statistics other than the gradient is used \cite{bigdata2,cesa_book,optimization}. Even though gradient based methods suffer from this convergence issue, they are extensively used in big data applications due to such low computational demand \cite{optimizationbook}.\par 
Different from the gradient based algorithms, the well known second order Newton-Raphson methods, e.g, Online Newton Step, use the second order statistics, i.e., Hessian of the cost function \cite{Hazan}. Hence, they asymptotically achieve the performance of the "best" possible predictor much faster\cite{singer}. Existence of logarithmic regret bounds is theoretically guaranteed for this family of algorithms as well \cite{Hazan}. Additionally, the second order methods are robust and prone to highly varying data statistics, compared to the first order methods, since they keep track of the second order information \cite{singer,kozatsayed}. Therefore, in the sense of convergence rate and steady state error performances, Newton-Raphson methods considerably outperform the first order methods \cite{cesa_book,singer,optimization}. However, the second order methods offer a quadratic computational complexity, i.e., $O(M^2)$, while the gradient based algorithms provide a linear relation, i.e., $O(M)$. As a consequence, it is not usually feasible for real-life big data applications to utilize the merits of the second order algorithms \cite{optimizationbook}. \par
In this paper, we study sequential data prediction, where the consecutive feature vectors are the shifted versions of each other, i.e., for a feature vector of $\vec{x}_t = [x_t, x_{t-1}, ..., x_{t-M}]^T$, the upcoming vector is in the form of $\vec{x}_{t+1} = [x_{t+1}, x_t, ..., x_{t-M+1}]^T$. To this end, we introduce second order methods for this important problem with computational complexity only linear in the data dimension, i.e., $O(M)$. We achieve such an enormous reduction in computational complexity since there are only two entries changing from $\vec{x}_t$ to $\vec{x}_{t+1}$, where we avoid unnecessary calculations in each update. We do not use any statistical assumption on the data sequence other than the shifted nature of the feature vectors. Therefore, we present an approach that is highly appealing for big data applications since it provides the merits of the Newton-Raphson methods with a much lower computational cost.\par
Overall, in this paper, we introduce an online sequential data prediction algorithm that $i)$ processes only the currently available data without any storage, $ii)$ efficiently implements the Newton-Raphson methods, i.e., the second order methods $iii)$ outperforms the gradient based methods in terms of performance, $iv)$ has $O(M)$ computational complexity same as the first order methods and $v)$ requires no statistical assumptions on the data sequence. We illustrate the outstanding gains of our algorithm in terms of computational efficiency by using two sequential real life big datasets and compare the resulting error performances with the regular Newton-Raphson methods. 
\par

\section{Problem Description}\label{sec:prob}
In this paper, all vectors are real valued and column-vectors. We use lower case (upper case) boldface letters to represent vectors (matrices). The ordinary transpose is denoted as $\vec{x}^T$ for the vector $\vec{x}$. The identity matrix is represented by $\vec{I}_{M}$, where the subscript is used to indicate that the dimension is $M \times M$. We denote the $M$-dimensional zero vector as $\vec{0}_M$.\par
We study sequential data prediction, where we sequentially observe a real valued data sequence $\{x_t\}_{t \geq 0}$, $x_t \in \mathbbm{R}$. At each time $t$, after observing $\{x_t,x_{t-1},...,x_{t-M+1}\}$, we generate an estimate of the desired data, $\hat{x}_{t+1} \in \mathbbm{R}$, using a linear model as
\begin{equation}\label{linear_model}
	\hat{x}_{t+1} = \vec{w}_{t}^{T}\vec{x}_{t} + c_{t},
\end{equation}
where $\vec{x}_t \in \mathbbm{R}^M$ represents the feature vector of previous $M$ samples, i.e., $\vec{x}_t = [x_t,x_{t-1},...,x_{t-M+1}]^T$. Here, $\vec{w}_t \in \mathbbm{R}^M$ and $c_t \in \mathbbm{R}$ are the corresponding weight vector and the offset variable respectively at time $t$. With an abuse of notation, we combine the weight vector $\vec{w}_t$ with the offset variable $c_t$, and denote it by $\vec{w}_t = [\vec{w}_t ; c_t]$, yielding $\hat{x}_{t+1} = \vec{w}_{t}^{T}\vec{x}_{t}$, where $\vec{x}_t = [\vec{x}_t ; 1]$. As the performance criterion, we use the widely studied instantaneous absolute loss as our cost function, i.e., $\ell_t(\vec{w}_t) = \lVert e_t \rVert$, where the prediction error at each time instant is given by 
$ e_t = x_{t+1} - \hat{x}_{t+1}. $ \par
We adaptively learn the weight vector coefficients to asymptotically achieve the best possible fixed weight vector $\hat{\vec{w}}_n$, which minimizes the total prediction error after $n$ iteration, i.e.,  
\[\hat{\vec{w}}_n = \argmin_{\vec{w} \in \mathbbm{R}^M} \sum\limits_{t=0}^{n}\lVert x_{t+1} - \vec{w}^T\vec{x}_t\rVert,\]
for any $n$. The definition of $\hat{\vec{w}}_n$ is given for the absolute loss case. To this end, we use the second order Online Newton Step (ONS) algorithm to train the weight vectors. The ONS algorithm significantly outperforms the first order Online Gradient Descent (OGD) algorithm in terms of convergence rate and steady state error performance since it keeps track of the second order statistics of the data sequence \cite{cesa_book,Hazan,optimization}. The weight vector at each time is updated as
\begin{equation}\label{ONS}
\vec{w}_{t} = \vec{w}_{t-1} - \dfrac{1}{\mu}\vec{A}_{t}^{-1}\nabla_{t},
\end{equation}
where $\mu \in \mathbbm{R}$ is the step size and $\nabla_{t} \in \mathbbm{R}^M$ corresponds to the gradient of the cost function $\ell_t(\vec{w}_t)$ at time $t$ w.r.t. $\vec{w}_t$, i.e., $\nabla_{t} \triangleq\nabla \ell_{t}(\vec{w}_t)$. Here, the $M \times M$ dimensional matrix $\vec{A}_t$ is given by 
\begin{equation}\label{A_t}
\vec{A}_t = \sum\limits_{i=0}^{t}\nabla_i \nabla_i^T + \alpha \vec{I}_M,
\end{equation}   
where $\alpha > 0$ is chosen to guarantee that $\vec{A}_t$ is positive definite, i.e., $\vec{A}_t > 0$, and hence, invertible. Selection of the parameters $\mu$ and $\alpha$ is crucial for good performance \cite{Hazan}. Note that for the first order OGD algorithm, we have $\vec{A}_t = \vec{I}_M$ for all $t$, i.e., we do not use the second order statistics but only the gradient information. \par
Definition of $\vec{A}_t$ in (\ref{A_t}) has a recursive structure, i.e.,
$\vec{A}_t = \vec{A}_{t-1} + \nabla_t \nabla_t^T,$
with an initial value of $\vec{A}_{-1} = \alpha \vec{I}_M$.  
Hence, we get a straight update from $\vec{A}_{t-1}^{-1}$ to $\vec{A}_{t}^{-1}$ using the matrix inversion lemma \cite{sayedbook}
\begin{equation}\label{A_t inv update}
\vec{A}_t^{-1} = \vec{A}_{t-1}^{-1} - \dfrac{\vec{A}_{t-1}^{-1} \nabla_{t} \nabla_{t}^{T}\vec{A}_{t-1}^{-1}}{1 + \nabla_{t}^{T}\vec{A}_{t-1}^{-1}\nabla_{t}}.
\end{equation}
Multiplying both sides of (\ref{A_t inv update}) with $\nabla_t$ and inserting in (\ref{ONS}) yields
\begin{equation}\label{update}
\vec{w}_{t} = \vec{w}_{t-1} - \dfrac{1}{\mu} \bigg[\dfrac{\vec{A}_{t-1}^{-1}\nabla_t}{1 + \nabla_{t}^{T}\vec{A}_{t-1}^{-1}\nabla_{t}}\bigg].
\end{equation}\par
Although the second order update algorithms provide faster convergence rates and better steady state performances, computational complexity issue prohibits their usage in most real life applications \cite{optimization,sayedbook}. Since each update in (\ref{A_t inv update}) requires the multiplication of an $M \times M$ dimensional matrix with an $M$ dimensional vector for $\vec{x}_t \in \mathbbm{R}^M$, the computational complexity is in the order of $O(M^2)$, while the first order algorithms just need $O(M)$ multiplication and addition. As an example, in protein structure prediction, we have $M=1000$ deeming the second order methods 1000 times slower than the first order OGD algorithm \cite{protein}.\par
In the next section, we introduce a sequential prediction algorithm, which achieves the performance of the Newton-Raphson methods, while offering $O(M)$ computational complexity same as the first order methods.

\section{Efficient Implementation for Complexity Reduction} \label{algorithm} 
In this section, we construct an efficient implementation that is based on the low rank property of the update matrices. Instead of directly implementing the second order methods as in (\ref{A_t inv update}) and (\ref{update}), we use unitary and hyperbolic transformations to update the weight vector $\vec{w}_t$ and the inverse of the Hessian-related matrix $\vec{A}_t^{-1}$.  \par
We work on time series data sequences, which directly implies that the feature vectors $\vec{x}_t$ and $\vec{x}_{t+1}$ are highly related. More precisely, we have the following relation between these two consecutive vectors as
\begin{equation}\label{consecutive}
[x_{t+1}, \vec{x}_t^T] = [\vec{x}_{t+1}^T, x_{t-M+1}].
\end{equation}
This relation shows that consecutive data vectors carry quite the same information, which is the basis of our algorithm. We use the instantaneous absolute loss, which is defined as
\begin{equation}
\ell_t(\vec{w}_t) = \lVert x_{t+1}-\vec{w}_t^T\vec{x}_t \rVert.
\end{equation}
Although the absolute loss is widely used in the data prediction applications, it is not differentiable when $e_t = 0$. However, we resolve this issue by setting a threshold $\epsilon$ close to zero and not updating the weight vector when the absolute error is below this threshold, $\lVert e_t \rVert < \epsilon$. From (\ref{A_t inv update}) and (\ref{update}), the absolute loss results in the following update rules for $\vec{w}_t$ and $\vec{A}_t^{-1}$,
\begin{equation}\label{generic_update}
\vec{w}_{t} = \vec{w}_{t-1} \pm \dfrac{1}{\mu}\bigg[\dfrac{\vec{A}_{t-1}^{-1}\vec{x}_{t}}{1 + \vec{x}_{t}^{T}\vec{A}_{t-1}^{-1}\vec{x}_{t}}\bigg],
\end{equation}
\begin{equation}\label{generic_update_2}
\vec{A}_t^{-1} = \vec{A}_{t-1}^{-1} - \dfrac{\vec{A}_{t-1}^{-1}\vec{x}_{t} \vec{x}_{t}^{T}\vec{A}_{t-1}^{-1}}{1 + \vec{x}_{t}^{T}\vec{A}_{t-1}^{-1}\vec{x}_{t}},
\end{equation}
since $\nabla_t = \pm \vec{x}_t$ depending on the sign of the error. \par
It is clear that the complexity of the second order algorithms essentially results from the matrix-vector multiplication, $\vec{A}_{t-1}^{-1}\vec{x}_{t}$ as in (\ref{generic_update}). Rather than getting matrix $\vec{A}_{t-1}^{-1}$ from $\vec{A}_{t-2}^{-1}$ and then calculating the multiplication $\vec{A}_{t-1}^{-1}\vec{x}_{t}$ individually at each iteration, we develop a direct and compact update rule, which calculates $\vec{A}_{t-1}^{-1}\vec{x}_{t}$ from $\vec{A}_{t-2}^{-1}\vec{x}_{t-1}$ without any explicit knowledge of the $M \times M$ dimensional matrix $\vec{A}_{t-1}^{-1}$. \par
Similar to \cite{sayedbook}, we first define the normalization term of the update rule given in (\ref{generic_update}) as
\begin{equation}
\eta_t = 1 + \vec{x}_{t}^{T}\vec{A}_{t-1}^{-1}\vec{x}_{t}.
\end{equation}
Then, the difference between the consecutive terms $\eta_t$ and $\eta_{t-1}$ is given by
\begin{equation}\label{n_diff}
\eta_t - \eta_{t-1} = \vec{x}_{t}^{T}\vec{A}_{t-1}^{-1}\vec{x}_{t} - \vec{x}_{t-1}^{T}\vec{A}_{t-2}^{-1}\vec{x}_{t-1}.
\end{equation}
We define the $(M+1) \times 1$ dimensional extended vector $\tilde{\vec{x}}_{t} = [x_{t}, \vec{x}_{t-1}^T ]^T$ and get the following two equalities using the relation given in (\ref{consecutive}),
\begin{equation}\label{n_t-1}
	\eta_t
	= 1 + \tilde{\vec{x}}_{t}^T
	\begin{bmatrix}
	\vec{A}_{t-1}^{-1} & \vec{0}_{M} \\
	\vec{0}_{M}^T & 0
	\end{bmatrix}
	\tilde{\vec{x}}_{t},
\end{equation}

\begin{equation}\label{n_t-2}
	\eta_{t-1}
	= 1 + \tilde{\vec{x}}_{t}^T
	\begin{bmatrix}
	0 & \vec{0}_{M}^T \\
	\vec{0}_{M} & \vec{A}_{t-2}^{-1}
	\end{bmatrix}
	\tilde{\vec{x}}_{t}.
\end{equation}
Therefore, (\ref{n_diff}) becomes
\begin{equation}\label{n_t}
\eta_t - \eta_{t-1} = \tilde{\vec{x}}_{t}^T\Delta_{t-1}\tilde{\vec{x}}_{t},
\end{equation}
where the update term $\Delta_{t-1}$ is defined as
\begin{equation} \label{Delta}
\Delta_{t-1} \triangleq
\begin{bmatrix}
\vec{A}_{t-1}^{-1} & \vec{0}_{M} \\
\vec{0}_{M}^T & 0
\end{bmatrix}
-
\begin{bmatrix}
0 & \vec{0}_{M}^T \\
\vec{0}_{M} & \vec{A}_{t-2}^{-1}
\end{bmatrix}.
\end{equation}
%
This equation implies that we do not need the exact values of $\vec{A}_{t-1}^{-1}$ and $\vec{A}_{t-2}^{-1}$ individually and it is sufficient to know the value of the defined difference $\Delta_{t-1}$ for the calculation of $\eta_t$. Moreover, we observe that the update term can be expressed in terms of rank 2 matrices, which is the key point for the reduction of complexity. \par
Initially, we assume that $x_{t} = 0$ for $t < 0$, which directly implies $\vec{A}_{-1}^{-1} = \vec{A}_{-2}^{-1} = \frac{1}{\alpha}\vec{I}_{M}$ using (\ref{A_t}). Therefore, $\Delta_{-1}$ is found as  
\begin{equation}
\Delta_{-1} = \dfrac{1}{\alpha}\text{ diag}\{1,0,\dots,0,-1\}.
\end{equation}
At this point, we define the $(M+1) \times 2$ dimensional matrix $\Lambda_{-1}$ and the $2 \times 2$ dimensional matrix $\Pi_{-1}$ as 
\begin{equation}
\Lambda_{-1} = \sqrt{\dfrac{1}{\alpha}}
	\begin{bmatrix}
		1 & 0 & \dots & 0 & 0 \\
		0 & 0 & \dots & 0 & 1 \\
	\end{bmatrix}^T,
	\Pi_{-1} = 
	\begin{bmatrix}
		1 & 0 \\
		0 & -1 \\
	\end{bmatrix},
\end{equation}
to achieve the equality given by
\begin{equation}
\Delta_{-1} = \Lambda_{-1}\Pi_{-1}\Lambda_{-1}^T.
\end{equation}
We show, at the end of the discussion, that once the rank 2 property is achieved, it holds for all $t \geq 0$. By using the reformulation of the difference term, we restate the $\eta_t$ term given in (\ref{n_t}) as 
\begin{equation}\label{nu_t}
\eta_t = \eta_{t-1} + \tilde{\vec{x}}_{t}^T\Lambda_{t-1}\Pi_{t-1}\Lambda_{t-1}^T\tilde{\vec{x}}_{t}.
\end{equation}
For the further discussion, we prefer matrix notation and represent (\ref{nu_t}) as 
\begin{equation}
	\begin{bmatrix}
		\sqrt{\eta_t} & \vec{0}_2^T \\
	\end{bmatrix}
	\begin{bmatrix}
		\sqrt{\eta_t} \\
		\vec{0}_2 \\
	\end{bmatrix}
	=
	\begin{bmatrix}
		\sqrt{\eta_{t-1}} & \tilde{\vec{x}}_{t}^T\Lambda_{t-1} \\
	\end{bmatrix}
	\Theta_{t-1}
	\begin{bmatrix}
		\sqrt{\eta_{t-1}} \\
		\Lambda_{t-1}^T\tilde{\vec{x}}_{t} \\
	\end{bmatrix},
\end{equation}
where $\Theta_{t-1}$ is defined as 
\begin{equation}
	\Theta_{t-1} \triangleq	
	\begin{bmatrix}
		1 & \vec{0}_2^T \\
		\vec{0}_2 & \Pi_{t-1} \\
	\end{bmatrix}.
\end{equation}
We first employ a unitary Givens transformation $\vec{H}_{G,t}$ in order to zero out the second element of the vector $[\sqrt{\eta_{t-1}}, \tilde{\vec{x}}_t^T \Lambda_{t-1}]$ and then use a $\Theta_{t-1}$-unitary Hyperbolic rotation $\vec{H}_{HB}$, i.e., $\vec{H}_{HB,t}\Theta_{t-1}\vec{H}_{HB,t}^T = \Theta_{t-1}$, to eliminate the last term \cite{sayedbook2}. Consequently, we achieve the following update rule
\begin{equation}
	\begin{bmatrix}
		\sqrt{\eta_t} & \vec{0}_2^T \\ 
	\end{bmatrix}
	= 
	\begin{bmatrix}
		\sqrt{\eta_{t-1}} & \tilde{\vec{x}}_{t}^T\Lambda_{t-1} \\
	\end{bmatrix}
	\vec{H}_t,
\end{equation}
where $\vec{H}_t$ represents the overall transformation process. Existence of these transformation matrices is guaranteed \cite{sayedbook}. This update gives the next normalization term $\eta_t$, however, for the $(t+1)^{th}$ update, we also need the updated value of $\Lambda_{t-1}$, i.e., $\Lambda_t$, explicitly. Moreover, even calculating the $\Lambda_t$ term is not sufficient, since we also need the individual value of the vector $\vec{A}_{t-1}^{-1}\vec{x}_t$ to update the weight vector coefficients. \par
We achieve the following equalities based on the same argument that we used to get (\ref{n_t-1}) and (\ref{n_t-2})
\begin{equation}\label{R_t-1}
	\begin{bmatrix}
		\vec{A}_{t-1}^{-1}\vec{x}_{t} \\
		0 
	\end{bmatrix}
	=
	\begin{bmatrix}
		\vec{A}_{t-1}^{-1} & \vec{0}_{M} \\
		\vec{0}_{M}^T & 0
	\end{bmatrix}
	\tilde{\vec{x}}_{t},
\end{equation}

\begin{equation}\label{R_t-2}
	\begin{bmatrix}
		0 \\
		\vec{A}_{t-2}^{-1}\vec{x}_{t-1}  
	\end{bmatrix}
	=
	\begin{bmatrix}
	0 & \vec{0}_{M}^T \\
	\vec{0}_{M} & \vec{A}_{t-2}^{-1}
	\end{bmatrix}
	\tilde{\vec{x}}_{t}.
\end{equation}
Here, by subtracting these two equations, we get
\begin{equation}\label{Rx}
	\begin{bmatrix}
		\vec{A}_{t-1}^{-1}\vec{x}_{t} \\
		0 
	\end{bmatrix}
	= 
	\begin{bmatrix}
		0 \\
		\vec{A}_{t-2}^{-1}\vec{x}_{t-1}  
	\end{bmatrix}
	+ \Delta_{t-1}\tilde{\vec{x}}_{t}.
\end{equation}

We emphasize that the same transformation $\vec{H}_t$, which we used to get $\sqrt{\eta_t}$, also transforms $\Lambda_{t-1}$ to $\Lambda_{t}$ and $\vec{A}_{t-2}^{-1}\vec{x}_{t-1}$ to $\vec{A}_{t-1}^{-1}\vec{x}_{t}$, if we extend the transformed vector as follows
\begin{equation}\label{transform}
	\begin{bmatrix}
		\sqrt{\eta_{t-1}} & \tilde{\vec{x}}_{t}^T\Lambda_{t-1} \\
		\dfrac{1}{\sqrt{\eta_{t-1}}}
		\begin{bmatrix}
			0 \\
			\vec{A}_{t-2}^{-1}\vec{x}_{t-1} \\
		\end{bmatrix}
		& \Lambda_{t-1} \\
	\end{bmatrix}
	\vec{H}_t
	= 
	\begin{bmatrix}
		\sqrt{\eta_t} & \vec{0}_2^T \\
		\vec{q} & \vec{Q} \\
	\end{bmatrix},
\end{equation}
where we show that $\vec{q} = \frac{1}{\sqrt{\eta_{t}}} [\vec{x}_t^T\vec{A}_{t-1}^{-1}, 0]^T$ and $\vec{Q} = \Lambda_t$. We denote (\ref{transform}) as $\vec{B}\vec{H}_t = \tilde{\vec{B}}$, where $\vec{B}$ represents the input matrix and $\tilde{\vec{B}}$ states the output matrix of the transformation. Then, the following equality is achieved 
\begin{equation}\label{trans_eq}
	\tilde{\vec{B}}\Theta_{t-1}\tilde{\vec{B}}^T = \vec{B}\Theta_{t-1}\vec{B}^T
\end{equation} 	  
since $\vec{H}_t$ is $\Theta_{t-1}$ unitary, i.e., $\vec{B}\vec{H}_t\Theta_{t-1}\vec{H}_t^T\vec{B}^T  = \vec{B}\Theta_{t-1}\vec{B}^T$.
Equating the elements of matrices in both sides of (\ref{trans_eq}) yields
\begin{equation}\label{q_Q}
	\begin{split}
		\vec{q}\sqrt{\eta_t} &= 
		\begin{bmatrix}
			0 \\
			\vec{A}_{t-2}^{-1}\vec{x}_{t-1} \\
		\end{bmatrix} + 
		\Delta_{t-1}\tilde{\vec{x}}_{t}, \\
		\vec{q}\vec{q}^T + \vec{Q}\Pi_{t-1}\vec{Q}^T &=
		\frac{1}{\eta_{t-1}}
		\begin{bmatrix}
			0 \\
			\vec{A}_{t-2}^{-1}\vec{x}_{t-1} \\
		\end{bmatrix}
		\begin{bmatrix}
			0 \\
			\vec{A}_{t-2}^{-1}\vec{x}_{t-1} \\
		\end{bmatrix}^T
		+ \Delta_{t-1}. \\
	\end{split}
\end{equation}
We know from (\ref{Rx}) that the left hand side of the first term in (\ref{q_Q}) equals to $[\vec{x}_t^T\vec{A}_{t-1}^{-1}, 0]^T$ and $\vec{q}$ is given by
\begin{equation}\label{q}
	\vec{q} = \dfrac{1}{\sqrt{\eta_t}}
	\begin{bmatrix}
		\vec{A}_{t-1}^{-1}\vec{x}_{t} \\
		0 \\
	\end{bmatrix}.
\end{equation}
Hence, we identify the value of $\vec{Q}$ matrix using the second term in (\ref{q_Q}) as 
\begin{equation}
	\begin{split}
		\vec{Q}\Pi_{t-1}\vec{Q}^T &= 
		\begin{bmatrix}
			0 & \vec{0}_M^T \\
			\vec{0}_M & \dfrac{\vec{A}_{t-2}^{-1}\vec{x}_{t-1}\vec{x}_{t-1}^T\vec{A}_{t-2}^{-1}}{\eta_{t-1}} \\
		\end{bmatrix}\\
		&+\bigg(
		\begin{bmatrix}
			\vec{A}_{t-1}^{-1} & \vec{0}_M \\
			\vec{0}_M^T & 0 \\
		\end{bmatrix}
		-
		\begin{bmatrix}
			0 & \vec{0}_M \\
			\vec{0}_M^T & \vec{A}_{t-2}^{-1} \\
		\end{bmatrix}\bigg)\\
		&- \vec{q}\vec{q}^T,
	\end{split}
\end{equation}
where we expand the $\Delta_{t-1}$ term using its definition given in (\ref{Delta}). We know that the term $\frac{1}{\eta_{t-1}}\vec{A}_{t-2}^{-1}\vec{x}_{t-1}\vec{x}_{t-1}^T\vec{A}_{t-2}^{-1}$ equals to the difference $\vec{A}_{t-2}^{-1} - \vec{A}_{t-1}^{-1}$ using the update relation (\ref{generic_update_2}). Therefore, substituting this equality and inserting the value of $\vec{q}$ yields
\begin{equation}
	\begin{split}
		\vec{Q}\Pi_{t-1}\vec{Q}^T &= \bigg(
		\begin{bmatrix}
			0 & \vec{0}_M \\
			\vec{0}_M^T & \vec{A}_{t-2}^{-1} \\
		\end{bmatrix}
		-
		\begin{bmatrix}
			0 & \vec{0}_M \\
			\vec{0}_M^T & \vec{A}_{t-1}^{-1} \\
		\end{bmatrix}\bigg)\\
		&+\bigg(
		\begin{bmatrix}
			\vec{A}_{t-1}^{-1} & \vec{0}_M \\
			\vec{0}_M^T & 0 \\
		\end{bmatrix}
		-
		\begin{bmatrix}
			0 & \vec{0}_M \\
			\vec{0}_M^T & \vec{A}_{t-2}^{-1} \\
		\end{bmatrix}\bigg)\\
		&-\bigg(
		\begin{bmatrix}
		\vec{A}_{t-1}^{-1} & \vec{0}_M \\
		\vec{0}_M^T & 0 \\
		\end{bmatrix}
		-
		\begin{bmatrix}
		\vec{A}_{t}^{-1} & \vec{0}_M \\
		\vec{0}_M^T & 0 \\
		\end{bmatrix}\bigg)\\
		&=
		\begin{bmatrix}
			\vec{A}_{t}^{-1} & \vec{0}_M \\
			\vec{0}_M^T & 0 \\
		\end{bmatrix}
		-
		\begin{bmatrix}
			0 & \vec{0}_M \\
			\vec{0}_M^T & \vec{A}_{t-1}^{-1} \\
		\end{bmatrix} \\
		&= \Delta_{t} \\
		&= \Lambda_{t}\Pi_{t}\Lambda_{t}^T.
	\end{split}
\end{equation}
This equality implies that $\Pi$ is time invariant, i.e., $\Pi_{t-1} = \Pi_{t}$ and $\vec{Q}$ is given as
\begin{equation}
	\vec{Q} = \Lambda_{t}.
\end{equation}
Hence, we show that when the low rank property of the difference term $\Delta_t$ is achieved for $t = i-1$, it is preserved for the iteration $t = i$, for $i \geq 0$. Therefore, the transformation in (\ref{transform}) gives all the necessary information and provides a complete update rule. As a result, the weight vector is updated as 
\begin{equation}
	\vec{w}_{t} = 
	\begin{cases}
		\vec{w}_{t-1} + \text{sgn}(e_t)\dfrac{1}{\mu}\bigg[\dfrac{\vec{A}_{t-1}^{-1}\vec{x}_{t}}{\eta_t} \bigg] , \text{ if } \lVert e_t \rVert > \epsilon\\
		\vec{w}_{t-1}, \text{ otherwise }
	\end{cases},
\end{equation}
where the individual value of $\vec{A}_{t-1}^{-1}\vec{x}_{t}$ is found by multiplying (\ref{q}) by $\sqrt{\eta_t}$, which is the left upper most entry of the transformed matrix $\tilde{\vec{B}}$, and taking the first $M$ elements. The complete algorithm is provided in Algorithm 1 with all initializations and required updates. \par
The processed matrix $\vec{B}$ has the dimensions $(M+2) \times 3$, which results in the computational complexity of $O(M)$. Since there is no statistical assumptions, we obtain the same error rates compared to the regular implementation. 
\begin{algorithm}[t]
	\KwData{$\{\vec{x}_t\}_{t \geq 0}$ sequence}
	Choose $\alpha > 0$, window size $M$ and the step size $\mu$ \;
	$\Lambda_{-1} = \sqrt{\dfrac{1}{\alpha}}
	\begin{bmatrix}
	1 & 0 & \dots & 0 & 0 \\
	0 & 0 & \dots & 0 & 1 \\
	\end{bmatrix}^T$ \;
	$\Pi = 
	\begin{bmatrix}
	1 & 0 \\
	0 & -1 \\
	\end{bmatrix}$,
	$\Theta = 	
	\begin{bmatrix}
	1 & \vec{0}_2^T \\
	\vec{0}_2 & \Pi \\
	\end{bmatrix}$ \;
	$\vec{x}_0 = \vec{0}_M$, 
	$\vec{w}_{0} = \vec{0}_M$, 
	$\eta_{-1} = 1$, 
	$\vec{\rho}_{-1} = \vec{0}_M$ \;
	\While{$t \geq 0$}{
		$\tilde{\vec{x}}_t = [x_t, \vec{x}_{t}^T]^T$\;
		$\hat{x}_{t+1} = \vec{w}_{t}^T\vec{x}_t$\;
		$e_t = x_{t+1} - \hat{x}_{t+1}$\;
		$\vec{B} =	
		\begin{bmatrix}
		\sqrt{\eta_{t-1}} & \tilde{\vec{x}}_{t}^T\Lambda_{t-1} \\
		\begin{bmatrix}
		0 \\
		\vec{\rho}_{t-1} \\
		\end{bmatrix}
		& \Lambda_{t-1} \\
		\end{bmatrix}$\;
		Determine a Givens rotation $\vec{H}_{G,t}$ for $\vec{B}$\;
		$\acute{\vec{B}} = \vec{B}\vec{H}_{G,t}$\;
		Determine a Hyperbolic rotation $\vec{H}_{HB,t}$ for $\acute{\vec{B}}$\;
		$
		\begin{bmatrix}
		\sqrt{\eta_t} & \vec{0}_2^T \\
		\begin{bmatrix}
		\vec{\rho}_t \\
		0 \\
		\end{bmatrix} & \Lambda_{t} \\
		\end{bmatrix} = \acute{\vec{B}}\vec{H}_{HB,t}$\;
		\uIf{$\lVert e_t \rVert > \epsilon$}{
			$\vec{w}_{t+1} = \vec{w}_{t} + \text{sgn}(e_t) \dfrac{1}{\mu}\bigg[\dfrac{\vec{\rho}_{t}\sqrt{\eta_t}}{\eta_t}\bigg]$ \;
		}
		\Else{
			$\vec{w}_{t+1} = \vec{w}_t$\;
		}
		$\vec{x}_t = [x_t, x_{t-1}, \dots , x_{t-M+1}]^T$\;
	}
	\caption{Fast Online Newton Step}
\end{algorithm} 
\section{Simulations}\label{simulations}
In this section, we illustrate the efficiency of our algorithm on real life sequential big datasets. We implement both the regular and the fast ONS algorithms on two different datasets, one of which is the CMU ARCTIC speech dataset where a professional US English male speaker reads 1132 distinct utterances\cite{speech}. The recording is performed at 16 KHz and there exist more than 50 million samples. The second dataset is a weather forecasting trace provided by Davis Weather station in Amherst, Massachusetts\cite{umass}. We also implement the first order OGD algorithm on this dataset to demonstrate the performance comparison with the second order methods in terms of computational efficiency and convergence rates. Temperature data was collected every 5 minutes during a period of 7 years from 2006 to 2013. There exist more than 600 thousand samples. Hence, both datasets are suitable for simulating big data scenarios. The data sequences are scaled to the range $[-1,1]$.\par  
\subsection{Computational Complexity Analysis}
\begin{figure}[t]
	\includegraphics[scale=0.6]{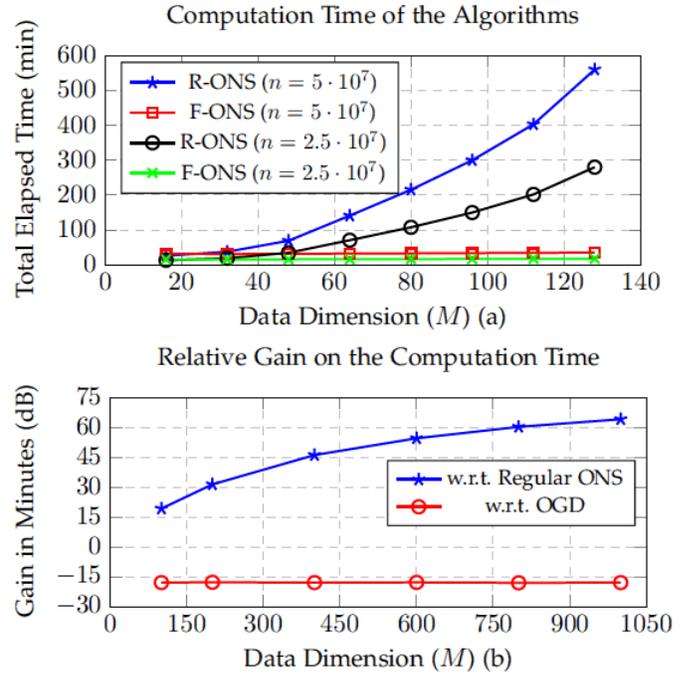}
	\caption{(a) Comparison of the computation time. R-ONS: Regular ONS, F-ONS: Fast ONS (b) Relative gain on the computation time with respect to the Regular ONS and the OGD algorithms when the Fast ONS algorithm is used.}
	\label{Speed}
\end{figure}

As the first experiment, we examine the computation time of both the proposed efficient ONS algorithm and the regular ONS algorithm. We first work on the two partitions of CMU ARCTIC speech dataset with lengths $n = 5 \cdot 10^7$ and $n = 2.5 \cdot 10^7$, and measure the corresponding total processing time. Sequences of different lengths are chosen to illustrate the effect of increasing data length. For both sequences, we choose feature vectors, $\vec{x}_t \in \mathbbm{R}^M$, with several dimensions ranging from $M=16$ to $M=128$. In Fig. \ref{Speed}.a, we demonstrate the computation time comparisons of the regular and the fast implementations of ONS algorithm. As expected from our results, complexity of the regularly implemented ONS algorithm shows a quadratic relation with the dimension of the feature vectors, $M$. However, the proposed efficient implementation provides a linear relation. A substantial observation from Fig. \ref{Speed}.a is that, with an increasing dimensionality of the space of feature vectors, the reduction in the complexity becomes outstanding. We also observe that the growth in the dataset length causes the same linear effect on both algorithms, i.e., doubling the total length $n$ results in the doubled computation time. \par
We then consider the weather forecasting temperature dataset, where in this case total length of the data sequence is not as much as the previous dataset. Therefore, we specifically concentrate on much larger values for the dimension of the feature vectors. Here, we choose the dimension of the space of feature vectors ranging from $M = 100$ to $M = 1000$ and total length of the data sequence is $n = 6 \cdot 10^5$. In Fig. \ref{Speed}.b, we illustrate the relative gain of the introduced fast ONS algorithm with respect to the regular implementation of ONS and the OGD algorithm in terms of total computation time. We observe that the relative computation time gain of the presented algorithm shows a significant improvement in comparison with the regular ONS, as the data dimension increases. However, we also observe that the relative gain falls into the negative region when compared with the first order OGD algorithm. This is an expected result, since the OGD uses only an $M$ dimensional vector in each iteration, whereas the fast ONS uses an $(M+2) \times 3$ dimensional matrix and performs additional transformation operations to update the weight vectors. Besides, the negative gain remains constant, since both algorithms eventually have the complexity in the same order of $O(M)$.  
\par

\subsection{Numerical Stability Analysis}
\begin{figure}[t]
	\includegraphics[scale=0.6]{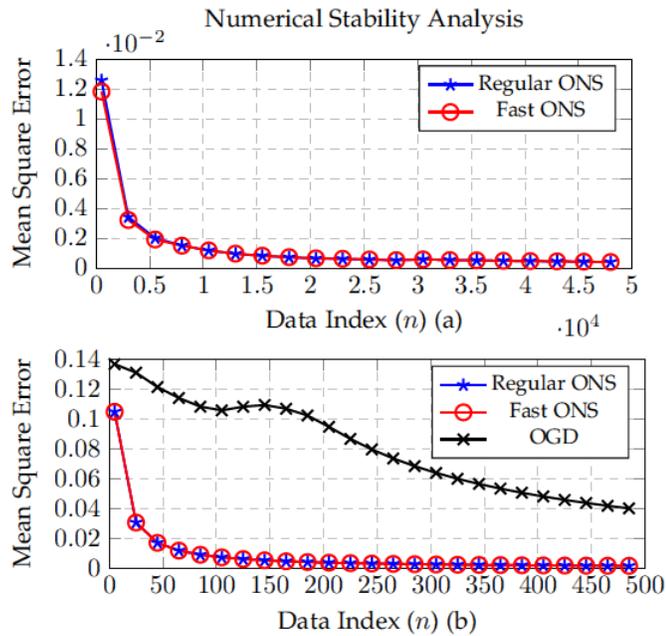}
	\caption{(a) Data dimension: $M = 64$ for both algorithms. (b) Data dimension: $M = 400$ for all algorithms.}
	\label{MSE_wheather}
\end{figure}

We theoretically show that the introduced algorithm efficiently implements the ONS algorithm without any statistical assumptions or any information loss. Hence, both the regular and the fast ONS algorithms offer the same error performances. However, there might occur negligible numerical differences as a consequence of the finite precision of real life computing systems. In the second part of the experiments, we examine the effects of numerical calculations on the mean square error curves of both the regular and the fast ONS algorithms. We first consider the CMU ARCTIC speech dataset with $n = 5 \cdot 10^4$ samples since we observe that the algorithms reach the steady state for this $n$. The dimension of the feature vectors is chosen as $M=64$, and the learning rates are determined as 0.003 for both algorithms. We demonstrate the comparison of mean square error curves in Fig. \ref{MSE_wheather}.a. A direct and significant observation is that the efficient implementation is numerically stable. There is no observable difference between the mean square error curves in terms of both convergence and steady state. \par

We work on the temperature tracking dataset as well for the numerical stability analysis. In this case, we include the OGD algorithm into the comparison and increase the dimension of feature vectors to $M = 400$ for all algorithms. Here, we examine the effect of large dimensionality on the numerical performance of the proposed algorithm and also compare the error performances with the first order OGD algorithm. Similar to the first case, we only represent the first 500 samples since the second order algorithms reach the steady state for this point. The learning rates are set to 0.001 for the regular and the fast ONS and 0.1 for the OGD algorithm. In Fig. \ref{MSE_wheather}.b, we illustrate the corresponding mean square error curves. Same as the previous analysis, the fast ONS algorithm shows numerically no difference compared to the regular ONS algorithm. Hence, even for such high dimensional feature vectors, the proposed algorithm remains numerically stable. Additionally, we illustrate that the first order OGD algorithm shows less than adequate performance in terms of convergence rate. Therefore, negative gain on the computation time observed in the previous experiment becomes insignificant when we consider the mean square error analysis in Fig. \ref{MSE_wheather}.b. \par

\section{Conclusion}
In this paper, we investigate online sequential data prediction problem for high dimensional data sequences. Even though the second order Newton-Raphson methods achieve superior performance, compared to the gradient based algorithms, the problem of extremely high computational cost prohibits their usage in real life big data applications. For an $M$ dimensional feature vector, the computational complexity of these methods increases in the order of $O(M^2)$. To this end, we introduce a highly efficient implementation that reduces the computational complexity of the Newton-Raphson methods from $O(M^2)$ to $O(M)$. The presented algorithm does not require any statistical assumption on the data sequence. We only use the similarity between the consecutive feature vectors without any information loss. Hence, our algorithm offers the outstanding performance of the second order methods with the low computational cost of the first order methods. We illustrate that the efficient implementation of Newton-Raphson methods attains significant computational gains, as the data dimension grows. We also show that our algorithm is numerically stable.

\bibliographystyle{IEEEtran}
\bibliography{IEEEabrv,double_bib}

\end{document}